\documentclass[prb,preprint]{revtex4-1} 

\usepackage{amsmath}  
\usepackage{amsfonts} 
\usepackage{graphicx} 
\usepackage{epic,eepic} 
\begin{document}

\title{The Helmholtz' decomposition of decreasing and weakly increasing vector fields}

\author{D. Petrascheck}
\author{R. Folk}

\affiliation{Institute for Theoretical Physics\\University Linz\\Altenbergerstr. 69, 
Linz, Austria}

\begin{abstract}
Helmholtz' decomposition theorem for vector fields is presented usually with too strong restrictions on the fields.
Based on the work of Blumenthal of 1905 it is shown that the decomposition of vector fields is not only possible
for asymptotically weakly decreasing vector fields, but even for vector fields, which asymptotically increase sublinearly.
Use is made of a regularization of the Green's function and 
the mathematics of the proof is formulated as simply as possible.
We also show a few examples for the decomposition of vector fields including the electric dipole radiation.
\end{abstract}
\maketitle

\section{Introduction}  

According to the Helmholtz' theorem one can decompose a given vector field $\vec v(\vec x)$
into a sum of two vector fields $\vec v_l(\vec x)$ and $\vec v_t(\vec x)$ where $\vec v_l$ is irrotational (curl-free)
and $\vec v_t$ solenoidal (divergence-free), if the vector field  fulfills certain conditions on continuity and
asymptotic decrease
($r\to \infty$). 
Here $\vec x$ is the position vector in three-dimensional space and $r=|\vec x|$ its absolute value. 

The two parts of the vector field can be expressed as gradient of a scalar potential and curl of a vector potential, respectively.
Concerning the validity, the uniqueness of the decomposition and the existence of the respective potentials one finds
different conditions.
 
The fundamental theorem for vector fields is historically based on Helmholtz' work on vortices \cite{helmholtz,stokes}
and therefore also known as Helmholtz' decomposition theorem.
For hydrodynamics  this theorem is of particular relevance, since the fluid fields of the decomposition have
the physical properties of freedom of vorticity and incompressibility, which for each field makes the analysis simpler.
Especially for the visualization of vector fields the decomposition theorem is of importance \cite{bhatia}.
 
F\"oppl \cite{abraham} introduced the decomposition theorem into electrodynamics.
He assumed a finite extension of the sources and vortices and therefore assumed a behavior for the corresponding
vector field of the form $|\vec v| \sim 1/r^2$ for  $|\vec x|=r \to \infty$.
However, his proof allows less restrictive conditions, namely an asymptotic decay of the field only somewhat stronger than $1/r$.
The decomposition theorem can be found in one of these formulations in most textbooks or lecture notes on electrodynamics.

Already in 1905 Otto Blumenthal \cite{blumenthal} proved, that any vector field, that goes to zero asymptotically
can be decomposed in a curl-free and a divergence-free part (weak version).
His formulation reads as follows \cite{zitat}:

''Let $\vec v$ be a vector, which is in addition to arbitrary many derivatives everywhere finite and continuous
and vanishes at infinity with its derivatives; then one can decompose this vector always  into two vectors,
a curl-free $\vec v_l$ and a divergence-free $\vec v_t$, such that
\begin{align} \tag{$^*$} \label{theorem1}
\vec v = \vec v_l (\vec x) + \vec v_t (\vec x) \, .
\end{align}
The vectors $\vec v_l$ and $\vec v_t$ diverge asymptotically weaker than $\ln r$.\\
In addition one has  the following proposition for uniqueness:
$\vec v_l$ and $\vec v_t$ are unique up to an additive constant vector, because of  
the given properties.''  No further specification for the behavior of the vector field was given.
\par \medskip \noindent 
This formulation was taken over in its essential statements by Sommerfeld \cite{sommerfeld} in 1944.
He noted further that the fundamental theorem of vector analysis, as he called it,
was already proven by Stokes \cite{stokes} in 1849 and in a more complete form by Helmholtz' paper of 1858.
 
The extension to a decay of $1/r$ and weaker is important for electromagnetic radiation but also for
a few configurations in electro- and magnetostatics.

Later on it was shown  that the conditions of continuity and differentiability can be weakened \cite{butzer,bhatia}
and that the theorem can be applied to vector fields behaving according to a certain power law \cite{tran-cong}.
Based on Blumenthal's method of regularization of the Green's function Neudert and Wahl \cite{wahl} investigated
among other things the asymptotic behavior of a vector field $\vec v$ if its sources div $\vec v$ and vortices curl $\vec v$
fulfill some conditions including differentiability and asymptotic decay. 

These developments remained to a large extent unnoticed in the physical literature \cite{panofsky} and
in mathematical physics \cite{gross}.
Thus it was necessary to show the validity of the decomposition theorem for electromagnetic radiation fields\cite{strahlung}
that decay asymptotically with $1/r$ .

 First we develop a systematic method, the so called regularization method, which is the basis of Blumenthal's proof,
but is not explicated in its generality and its improvement  in order to be applicable to  vector fields, which decay asymptotically with a specified power law.
Then we reformulated the decomposition theorem including all potentials for such cases. 
It is shown how the levels of the regularization modifies the validity of the uniqueness for the  vector fields depending on their behavior at infinity.
The equations necessary to construct the decomposed parts are presented.
The necessity of a formulation of a proper asymptotic condition either for the irrotational or solenoidal part is pointed out. 

In the next section we apply the decomposition theorem to the electromagnetic dipole radiation making
clear why the traditional theorem is applicable and how the different quantities in the theorem are related to electrodynamics.

Then we formulate and proof the extension of the theorem and uniqueness up to a constant vector for sublinearly diverging vector fields.
A mathematical example without physical background is given in an Appendix. Such a case might be of interest for physics,
if one has sources (circulations) that remain finite even at infinity.  
In the conclusion we summarize the results.

\section{Regularization Method \label{met}} 
 
The solution $\phi_0(\vec x)$ of the Poisson equation
\begin{align} \label{poisson}
\Delta \phi_0 (\vec x)= -4\pi \rho(\vec x)
\end{align}
with the source density $\rho(\vec x)$ is found by introducing it's Green's function 
\begin{align}\label{g0}
G_0(\vec x,\vec x')&= \frac{1}{|\vec x' - \vec x|}  \\
\phi_0(\vec x)&=   \int d^3 x'\,\rho(\vec x')G_0(\vec x,\vec x') \, . 
\end{align}
If the solution exists in the whole domain of $\mathbb{R}^3$, the integral has to be finite. 
This is guaranteed by a sufficient decay of the integrand, either by a sufficient strong decay of the source density
and/or by a sufficient decrease of the Green's function.

In his work on the Helmholtz' theorem \cite{blumenthal} Blumenthal presented a method  to make this solution finite
(regularizing the solution) by changing  the Green's function of the Poisson equation, without changing the Poisson equation
(that means without changing the source density).  
Thus one can prove the existence of the potential for cases where the source density is less strong decreasing.  From this method it becomes clear how a systematic extension of the decomposition theorem  is possible.

Introducing an arbitrary point $\vec x_0$ (apart from the condition that $\rho(\vec x_0)$ is finite at this point;
regularization point) and noting that
$G_0(\vec x,\vec x')=G_0(\vec x  - \vec x_0,\vec x'-\vec x_0)$,
we expand $G_0$ in a power series in $\vec x -\vec x_0$
\begin{align} \label{exp}
G_0(\vec x,\vec x')&
 =\frac{1}{|\vec x'  - \vec x_0|}+\frac{(\vec x - \vec x_0) \cdot( \vec x'-\vec x_0)}{|\vec x'-\vec x_0|^3}
+ O\Big(\frac{1}{|\vec x'-\vec x_0|^3}\Big) \, .
\end{align}
A stronger  decrease for large $|\vec x'|$ of the Green's function is now reached by subtraction of the corresponding
expansion terms. 
We get the following set of stronger decreasing Green's functions
\begin{align}  \label{g1}
G_1(\vec x-\vec x_0,\vec x'-\vec x_0)&= G_0(\vec x,\vec x')- \frac{1}{|\vec x'  - \vec x_0|}
\\  \label{g2}
G_2(\vec x-\vec x_0,\vec x'-\vec x_0)&
=G_1(\vec x-\vec x_0,\vec x'-\vec x_0)-\frac{(\vec x - \vec x_0) \cdot( \vec x'-\vec x_0)}{|\vec x'-\vec x_0|^3}
\, .  
\end{align}
The asymptotic decrease of these modified Green's functions is as $\sim 1/r'^{1+i}$.
For $i \le 2$ the subtracted terms do not change the source density
\begin{align} \label{deltagi} 
\Delta G_i(\vec x -\vec x_0,\vec x'-\vec x_0) = - 4 \pi \delta(\vec x'- \vec x) &&
\mathrm{for}&& 0 \le i \le 2 \, .
\end{align}
But they allow to extend the range of the validity for which the existence of the potential (and the decomposition)
can be proven
\begin{align} \label{phierweitert}
\phi_i(\vec x)= \int d^3 x'\,\rho(\vec x')\,G_i(\vec x-\vec x_0,\vec x'-\vec x_0)
&& \mathrm{and} &&
\Delta \phi_i(\vec x) = - 4 \pi \rho(\vec x)\qquad \mathrm{for} \quad i \le 2
  \, .
\end{align}
The solutions $\phi_i(\vec x)$ differ only by a (divergence- and curl-free) solution of the Laplace equation,
i.e.
$\phi_0(\vec x)$ differs from $\phi_1(\vec x)$ by a constant value and from $\phi_2(\vec x)$ by a linear function,
both depending on $\vec x_0$.
\par \smallskip 
Trying to extend the range of validity even further one may
subtract the next (third) term in the expansion  \eqref{exp} from $G_2$ and obtains
\begin{align}  \label{g3}
G_3(\vec x-\vec x_0,\vec x'-\vec x_0)&= G_2(\vec x-\vec x_0,\vec x'-\vec x_0)
-\frac{1}{2}\big ((\vec x - \vec x_0)\cdot \vec \nabla'\big )^2\,\frac{1}{|\vec x'-\vec x_0|} .
\end{align}
But now $G_3$ fulfills the Poisson equation
\begin{align} 
\Delta G_3(\vec x -\vec x_0,\vec x'-\vec x_0) = - 4 \pi \big [ \delta(\vec x'- \vec x)
- \delta(\vec x'-\vec x_0) \big ]
\end{align}
 from which it follows, that $G_3$ leads to a solution of a modified Poisson equation
\begin{align} 
\Delta \phi_3(\vec x) = - 4 \pi \big [ \rho(\vec x) - \rho(\vec x_0)\big]
  \, .
\end{align}
Thus the method described here is not suitable for Green's functions $G_i$ with $i > 2$.
This means (as we will see later) that  vector fields which increase linearly or even stronger
will not be decomposed by the regularization method described here. 
                                                                            
Nevertheless one should remark that one can solve the Poisson equation even with $G_3$ if one 
subtracts the solution for the inhomogeneity $\rho(\vec x_0)$ 
\begin{align} \label{phi3}
\bar{\phi}_3(\vec x)=\int d^3 x'\,\rho(\vec x') G_3(\vec x - \vec x_0,\vec x'-\vec x_0) + 2\pi \frac{\rho(\vec x_0)}{3} \,r^2 .
\end{align}
We refer to this solution in section \ref{strongdiverging}. 

The relations 
\begin{align} \label{nabg2}
\vec \nabla G_{i+1}(\vec x-\vec x_0,\vec x'-\vec x_0) &= -\vec\nabla' G_i(\vec x-\vec x_0,\vec x'-\vec x_0)   
&\mathrm{for} \quad i\le 2
\end{align}
can  be derived from  \eqref{g1}, \eqref{g2} and \eqref{g3}.
They  are used a few times, mainly to compute the vector fields $\vec v_l$ and $\vec v_t$ and to establish relations between them. 

In the following we will restrict ourselves to the regularization point $\vec x_0=0$, because
the Green's functions are simpler without loss of generality. 
In this case the  potential is fixed to  $\phi(\vec x=0) = 0$. 
We will keep this choice in the remaining part of the paper as far as possible.
\section{The fundamental theorem of vector analysis}
As already  noticed, the formulation of the fundamental theorem rests in its form today on the work of Blumenthal.
However there are several reasons not to take the formulations of Blumenthal resp. Sommerfeld literally.
For instance the uniqueness of the decomposition into the fields of the sources and vortices,
was only shown up to a constant vector. 
We will formulate the conditions in such a form, that a strict uniqueness of the decomposition is given.
Furthermore in the proof, which will be given,  the potentials by which the decomposed fields are calculated,
are part of the theorem (strong version). 
It is common in electrodynamics to calculate the physical fields via the introduction of potentials.
Moreover since the proof of Blumenthal is somewhat complex and lengthy it is not found in detail in textbooks.
Therefore a shorter and more compact proof seems to be useful.
Thus we formulate the theorem in the following way:
\par \smallskip \noindent
{\sl Let $\vec v(\vec x)$ be an everywhere continuous differentiable vector field 
 with the asymptotic behavior
$\displaystyle \lim_{r \to \infty} v(r)\,r^\epsilon < \infty$, where $\epsilon > 0$}, 
{\sl then the decomposition 
\begin{equation}\label{vektorsatz}
\vec v(\vec x) = \vec v_l + \vec v_t = - \vec \nabla \phi(\vec x) + \vec \nabla \times \vec A(\vec x) 
\end{equation}
is unique with
\begin{align} \label{skalarpot} 
\phi(\vec x) &=   \frac{1}{4\pi}\int d^3 x'\,(\vec \nabla'\cdot \vec v(\vec x'))
\big (\frac{1}{|\vec x' - \vec x|}- \frac{1}{r'}\big )
\\ \label{vektorpot} 
\vec A(\vec x) &= \frac{1}{4\pi}\int d^3 x'\,(\vec \nabla' \times  \vec v(\vec x'))
\big (\frac{1}{|\vec x' - \vec x|}- \frac{1}{r'}\big ) .
\end{align}
}                                                                                      
\par \smallskip \noindent 
\textit{Remarks}:
\begin{itemize}

\item Curl- and divergence-free fields $\vec v_h$ can be added to $\vec v_l$ if they are subtracted from $\vec v_t$
without affecting the boundary conditions of $\vec v$.
Such harmonic vector fields are suppressed if one explicitely demands that $\vec v_l$ and/or $\vec v_t$ vanishes asymptotically
and establish a strict uniqueness of the decomposition.   
\item 
Usually the potentials $\phi(\vec x)$ and $\vec A(\vec x)$ are defined with the Green's function $G_0$
\eqref{g0}.
If they are finite, then there is no need for $G_1$ \eqref{g1}.   
However if the vector field $\vec v$ decays asymptotically as $1/r$ or weaker,  one generally has to 
use the Green's function $G_1$ as shown in  \eqref{skalarpot} and \eqref{vektorpot} 
in order to avoid divergences in the potentials $\phi(\vec x)$ and $\vec A(\vec x)$.
\item
As already mentioned in section \ref{met}, the potentials are fixed  to the values $\phi(0)=0$ and
$\vec  A(0)= 0$ by the choice of the regularization point 
$\vec x_0=0$.
This choice does not affect the vector fields $\vec v_l$ and $\vec v_t$. 
\item The vector potential $\vec A$ by its definition is purely transversal, $\vec \nabla \cdot \vec A=0$ (see proof below in App. \ref{vec25}).
\item   We want to stress the point that the decomposition theorem holds for any vector field independent of the type of physical equations the vector field might fulfill. On the other hand if one thinks of the electric 
field or the magnetic field as examples of the theorem, due to the Maxwell equations these fields are connected although with respect to the decomposition theorem they are independent.
\end{itemize}    
Let us define the source density $\rho(\vec x)$ and the vortex density $\vec j(\vec x)$ as
\begin{align}
\rho(\vec x) &= \frac{\vec \nabla \cdot \vec v(\vec x)}{4\pi} &
\vec j(\vec x) &= \frac{\vec \nabla \times \vec v(\vec x)}{4\pi},  
\end{align}
then the decomposition of the corresponding vector field in its irrotational (curl-free) and
solenoidal (divergence-free) parts leads to the result, that
\begin{align}
\vec \nabla \cdot \vec v_l(\vec x)&= 4\pi \rho(\vec x) &\mathrm{and} && \vec \nabla \times \vec v_l(\vec x)&= 0 
\\
\vec \nabla \times \vec v_t (\vec x) &= 4 \pi \vec j(\vec x) & \mathrm{and} && \vec \nabla \cdot \vec v_t(\vec x)&=0. 
\end{align}
\subsubsection{Proof of the fundamental theorem\label{phi21}}

First we show the existence of the scalar potential.
If the finiteness of the integral  \eqref{skalarpot} is proven, one gets  the field  $\vec v_l$ by calculating
the gradient of $\phi$. 
For this it is required that the integration and differentiation interchange.
Then one can show that $\vec \nabla \times \vec v_l=0$ and $\vec \nabla \cdot \vec v_l = \vec \nabla \cdot \vec v$. 

Subsequently one proceeds quite similarly for the vortex field by showing the existence of  \eqref{vektorpot} first,
then calculating  $\vec v_t$ and proving its properties  $\vec \nabla \times \vec v_t= \vec \nabla \times \vec v$
and $\vec \nabla \cdot \vec v_t=0$.
Finally we check that the sum $\vec v_l+\vec v_t=\vec v$.

If we show that the integral  \eqref{skalarpot} exists and is finite, then the longitudinal part $\vec v_l$ can be determined. 
We note, that the singularities  at $\vec x$ and at zero do not lead to a diverging contribution to the integral,  because of the antisymmetry of the integrand around the singularity.
More important is the asymptotic behavior of the integral for $r' \to \infty$.
We integrate then over the surface of a larger sphere with radius  $R$.
Now we have to take into account the regularization term  \eqref{g1}.
Since no assumptions have been made on the asymptotic behavior of the sources $\rho=\vec \nabla \cdot \vec v/4 \pi$
but only on  $\vec v$, we perform a partial integration.
This allows us to prove the convergence from the behavior of the vector field $\vec v$ alone.
Integrating over the volume of the sphere leads to 
\begin{align}
\phi (\vec x)  &\stackrel{R \gg r}{=}
\frac{1}{4\pi}\int_{S_R} d^3 x'\,(\vec \nabla' \cdot \vec v(\vec x'))\,
G_1(\vec x,\vec x') 
\\ \notag  &
= \frac{1}{4\pi}\oint_{\partial S_R}d \vec f' \cdot \underbrace{\vec v(\vec x')}
_{ \sim 1/R^\epsilon}
\underbrace{ G_1(\vec x,\vec x') }_{ \sim 1/R^{2}} 
-\frac{1}{4\pi}\int_{S_R} d^3 x'\,
\vec v(\vec x') \cdot \vec \nabla'
G_1(\vec x,\vec x') \, . 
\end{align}
The radius $R$ can be chosen in such a way, that  the field becomes small. 
Then the surface integral vanishes as $1/R^{\epsilon}$  and it remains to show convergence of the volume integral.

In order to achieve this  we separate the volume of integration into an inner volume of a sphere $S_R$
with radius  $R \gg r$ and the outer domain $r'  \ge R$
\begin{align} \label{phi}
\phi(\vec x)  &\stackrel{R \gg r}{=}
- \frac{1}{4\pi} \int_{S_R} d^3 x'\, \big (\vec v(\vec x') \cdot \vec \nabla' \big ) 
G_1(\vec x,\vec x') + \phi_a(\vec x).
\end{align}
The contribution of the outer domain to the potential has been indicated by $\phi_a(\vec x)$.
For an estimate of this term one can take the the Taylor expansion of $G_1$,  \eqref{g1},  and finds
\begin{align} \label{phia} 
|\phi_a(\vec x)| &= \Big |  \frac{-1}{4\pi} \int_{r' \ge R} d^3 x'\,
\big ( \vec v(\vec x')  \cdot \vec \nabla'\big )
\big ( \frac{\vec x  \cdot \vec x'}{r'^3} + ...\big )\Big |
\\ \notag &
\le \Big | \int_R ^\infty d r'\,r'^2\,v_0 \frac{1}{ r'^\epsilon} \big [ \frac{4r}{r'^3}+ O(\frac{1}{r'^4})\big ]\Big |
\approx 4 r v_0 \frac{1}{\epsilon R^\epsilon}.
\end{align} 
Thus the contribution of the outer domain to the potential vanishes  as  $1/\epsilon R^\epsilon$,
and the existence of $\phi(\vec x)$ has been proved. 


It should be proven that 
the negative gradient of $\phi$   \eqref{skalarpot} represents the curl-free part $\vec v_l$  of $\vec v$
\begin{align} \label{a.vl}
\vec v_l(\vec x) = - \vec \nabla \phi(\vec x) &= \frac{-1}{4\pi}\int d^3 x'\,\big (\vec \nabla'\cdot \vec v(\vec x')\big )
\,\vec \nabla G_1(\vec x,\vec x').
\end{align}
Since $\vec v_l$ is calculated from a potential curl\,$\vec v_l$ is zero.  
  
Now it should be shown that $\vec v_l$ has the same sources as $\vec v$ 
\begin{align}
\vec \nabla \cdot \vec v_l &= \frac{-1}{4\pi}\int d^3 x'\,
\big (\vec \nabla' \cdot \vec v(\vec x')\big )  \Delta G_1(\vec x, \vec x') 
= \vec \nabla \cdot  \vec v.   
\end{align}
Here we used the property   \eqref{deltagi} of the Green's function.
Both vector fields have indeed the same sources.
In  \eqref{a.vl} one can replace $\vec \nabla G_1(\vec x,\vec x')$ by $-\vec \nabla' G_0(\vec x,\vec x')$ and one obtains 
the longitudinal vector field in a manner that is known from the potential-theory
in electro- and magnetostatic
\begin{align} \label{a.vlb}
\vec v_l(\vec x)  &= \int d^3 x'\, \rho(\vec x')\,\vec \nabla' G_0(\vec x,\vec x').
\end{align}
The proof for the vector potential goes along the same lines and is shifted to appendix A.

\par \medskip

\subsubsection{Proof of uniqueness}
Now, the existence of the potentials  \eqref{skalarpot} and \eqref{vektorpot} has been proven.
In the last step of the proof, the uniqueness of the decomposition has to be shown. 

We have decomposed the vector field $\vec v$ in a source field $\vec v_l$ and a vortex field $\vec v_t$,
under the boundary condition that the total field $|\vec v|$ vanishes going to infinity.
In order to reach uniqueness of the decomposition we demand that 
$|\vec v_l|$ and in consequence also  $|\vec v_t|$  vanish going to infinity.  

Assume  two different decompositions of the vector field $\vec v  = \vec v_l  + \vec v_t= \vec v'_l  + \vec v'_t $ and consider the differences of the source and vortex fields. Then  the difference of the longitudinal vector field  $\vec v_d  = \vec v_l  - \vec v'_l$ is a divergence- and curl-free field that 
vanishes at infinity.
It can be derived  from a scalar potential $\phi_d$, which fulfills the Laplace-equation (harmonic function).
The only solution for the potential allowed  would be a constant
(see also the argumentation for \eqref{g2.laplacepot1} and \eqref{g2.flux}).
Thus the vector field $\vec v_d$ has to be zero and the decomposition is unique.
 
\section{Applications in electrodynamics}
\subsection{Static fields}

In electrodynamics the fundamental theorem of vector analysis is used especially (although not always mentioned)
in magneto-statics for magnetic fields in matter \cite{miller}.  We consider a permanent magnet, where no volumen current is present.
There (in the Gaussian system) the magnetization $4\pi  \vec M$ corresponds \cite{petra} to the vector field
$\vec v$ in the decomposition theorem, the magnetic field $-\vec H$ to the longitudinal (irrotational) part $\vec v_l$
and the magnetic induction $\vec B$ to the transversal (solenoidal) part $\vec v_t$. The source  is given by $\rho_H=-\vec\nabla\cdot \vec M$ and the circulation by $\vec j_H=  \vec\nabla\times\vec M$.

These fields are related by the material equation $4\pi \vec M= \vec B -\vec H$. This exactly corresponds to the decomposition theorem.
For such a case the sources and vortices are near the surface of the magnetic body, since the magnetization inside
is almost constant. 
In any case the sources and vortices are localized to a finite region and in consequence the corresponding source
and vortex field decay asymptotically at least as $1/r^2$. 
The total vector field $\vec v$ of the magnetization is zero outside the magnetic body. 

A quite similar situation occurs in electrostatics   in a medium with spontaneous polarization, where no free charges are present \cite{miller}.
There the vector field  $\vec v$ corresponds to the polarization $4\pi \vec P$, the source field  $\vec v_l$
to the electrostatic field $-\vec E$ and the vortex field  $\vec v_t$ to the dielectric displacement field  $\vec D$. Again the decomposition therem corresponds to the material equation $4\pi \vec P=\vec D-\vec E$.

Even in electro- and magnetostatics configurations with slow decreasing fields exist.
The electric field of an infinite  straight wire, which bears an electric charge,
decays as $\sim 1/\rho$, where $\rho$ is the distance to the wire.
If on the other hand the wire carries a current, then the magnetic field decays as $\sim 1/\rho$.
In both cases a regularization is appropriate to get the potentials from finite integrals over the sources.
 
\subsection{Time dependent fields: the electric field of an oscillating  dipole \label{rad}}
Periodically moved charge densities $\rho(\vec x,t)= \rho(\vec x)\,e^{- i \omega t}$ of frequency $\omega$ emit a
radiation field of the same frequency.
For simplicity we use the complex notation understanding the physical quantities (charge density, potential, fields)
always as the real parts of the corresponding complex quantities.
The radiation fields factorize in the same way as the sources $\vec v(\vec x,t)=\vec v(x) \,e^{-i \omega t}$,
where in $\vec v(\vec x)$  the dependence on the frequency $\omega$ resp. wave number $k=\omega/c$ has been suppressed.
A decomposition of the time independent vector field  $\vec v(\vec x)$ is possible, since the radiation field,
or more precisely its long range  part,
decays as $1/r$ and thus fulfills clearly the conditions of the decomposition theorem. 

If one starts from the assumption that the asymptotic behavior of the field has to be stronger than $1/r$,
additional considerations are needed in order to proof the decomposition of the radiation fields \cite{stewart}.
Radiation fields, which decay asymptotically as $1/r$ are rarely connected with the decomposition theorem.
One reason might be that in most of the textbooks on electrodynamics the result of Blumenthal's proof are not mentioned and
one gets the impression the decomposition theorem can only applied under additional conditions\cite{stewart}
as they are found in radiation fields like $e^{i kr}/r$.
The peculiarity of these cases is, that one does not need the regularization term, although one has a field of $O(1/r)$. 

Strictly speaking the conditions of the theorem are not fulfilled if the vector field has singularities
due to point sources.
This also holds for the radiation fields considered.
However the integration over the sources in  \eqref{skalarpot} and \eqref{vektorpot} remain finite.
The only consequence, in cases where a regularization is necessary, is that the regularization point
has to be different from the singular points due to the source.

The electric radiation field $\vec E(\vec x)\equiv \vec v(\vec x)$ 
of an oscillating point dipole $\vec p(t)=\vec p \, e^{- i \omega t} $  reads \cite{altzit} 
\begin{align} \label{g5.41} 
\vec v(\vec x)  = \frac{e^{i k r}}{r}\Big \{ k^2  \vec e_r  \times (\vec p  \times \vec e_r)
  + \frac{1}{r^2}(1  - i kr )\Big [ 3 (\vec p  \cdot \vec e_r) \vec e_r  -\vec p \Big ]\Big \} .\!   
\end{align}
 $\vec e_r = \vec x /r$ is the unit vector in the direction of $\vec x$ and
$\vec v(\vec x)$ is the spatial part of the electric field.  
For $k  = \omega/c=0$, one obtains of course the static dipole field.  
Let us first calculate the source and vortex density
\begin{align}
\rho_H (\vec x) &
=  \frac{\vec \nabla  \cdot \vec v}{4\pi}
=e^{i k r}(1  - i  k r )\rho_p(\vec x)\, \widehat{=}\,\rho_p(\vec x)
  & \rho_p(\vec x) &= - \vec p  \cdot \vec \nabla \delta(\vec x )
\\ \label{bfield}
\vec j_H(\vec x) &=  \frac{\vec \nabla  \times \vec v}{4\pi}
=-\frac{e^{i k r}}{4\pi}\,\frac{k^2}{r^2}(1- i k r)\,(\vec e_r  \times \vec p)=\frac{ik}{4\pi}\vec B(\vec x).    
\end{align}
$\rho_p(\vec x)$ is the localized charge density of the static dipole \cite{rho},
whereas the vortex density is extended in the whole domain decreasing for  $r \to \infty$ as the radiation field with $1/r$.
 It can be identified with the spatial part of the magnetic radiation field\cite{altzit} $\vec B$  apart from a factor, as expected from Faraday's law of induction.
Surprisingly  the wave number dependence in $\rho_H(\vec x)$, which in $\vec j_H(\vec x)$ comes from the retardation, drops out. 
This asymmetry has already been discussed by Brill and Goodman \cite{brill}. Hence the scalar potential is given by
\begin{align}
\phi_H(\vec x) &= \int d^3 x' \,\frac{\rho_H(\vec x')}{|\vec x'  - \vec x|} 
 = \frac{\vec p  \cdot \vec e_r}{r^2}=\phi_C(\vec x).
\end{align}
Multiplying by the factor $e^{-i \omega t}$ one obtains the quasi-static (acausal) dipole potential $\phi_C(\vec x, t)$
as it it known using the Coulomb gauge \cite{komm}.
One obtains the spatial part of the scalar potential in Coulomb gauge \cite{komm}, which the static (acausal) dipole potential.
From that it is clear that  the longitudinal decomposed vector field $\vec v_l$ is the quasi-static electric field of a point dipole
\begin{align} 
\vec v_l(\vec x) & = - \vec \nabla \phi_H(\vec x) = \big [ -\vec p  + 3 (\vec p  \cdot \vec e_r) \vec e_r \big ] \frac{1}{r^3}
\end{align}
and does not contribute to the electromagnetic radiation, which is  pure transversal. 
The decomposition is finally shown by calculating the transversal part  $\vec v_t= \vec \nabla \times \vec A_H$ according to
the theorem from of the vector potential
\begin{align} \label{ahele}
\vec A_H(\vec x) &= \int d^3 x' \,\frac{\vec j(\vec x')}{|\vec x'  - \vec x|} 
={k^2} \vec p  \times \vec e_r \big [
 \frac{e^{ i k r}}{i k r} + \frac{1}{ k^2r^2}  (e^{i k r} -1 )\big ]= \frac{i}{k}\vec B(\vec x) + \frac{\vec e_r}{r^2} \times \vec p.   
\end{align}
In electrodynamics one never defines a vector potential for the electric field, but it is known
from the Amp\`ere-Maxwell-equation that the electric field can be calculated via the curl of $\vec B$.
The longitudinal part of $\vec v_l$ is removed by the second term of $\vec A_H$, a quasistatic vector field.
Note that this is {\bf not} the vector potential $\vec A_C$ known from calculating the electric and magnetic fields in the Coulomb gauge
\begin{align} \label{vectorHC}
\vec A_C(\vec x) &=
-\frac{e^{i kr }}{r}\Big \{ i k \,  \vec e_r  \times (\vec p  \times \vec e_r)
- \big [ \vec p  - 3 (\vec e_r  \cdot \vec p) \vec e_r\big ] 
\big [\frac{1}{r} + \frac{i }{k r^2}(1  - e^{-i k r})\big ] \Big \} \notag \\
&=\frac{1}{ik}\big(\vec E(\vec x)+\nabla \phi_C(\vec x)\big).
\end{align}
Thus with the transverse field
\begin{align}
\vec v_t(\vec x) & = \frac{e^{i k r}}{r}\Big \{ k^2  \vec e_r  \times (\vec p  \times \vec e_r)
  + \frac{1}{r^2}(1  - i kr )\Big [ 3 (\vec p  \cdot \vec e_r) \vec e_r  -\vec p \Big ]\Big \}
-\frac{1}{r^2}\big [3 (\vec p \cdot \vec e_r) \vec e_r - \vec p]
\notag \\  &
= \vec E(\vec x) - \vec v_l(\vec x) .\!   
\end{align}
the causal character of the total electric radiation field $\vec v(\vec x)$ is restored \cite{rohrlich,jack}.

The same decomposition may be done for the magnetic radiation field, which however is trivial since the field is only transversal (see \eqref{bfield}). 
The vector potential fulfills  $\vec \nabla \cdot \vec A_H = 0$ and $\vec \nabla \times A_H=\vec B$, the same conditions as for the vector potential $\vec A_C$ in the Coloumb gauge. We obtain indeed
$\vec A_H(\vec x) = \vec A_C(\vec x)$.
The vortex density of the magnetic field is apart from a factor given by the same expression as the electric field
$\vec E(\vec x)$  \eqref{g5.41} of the electric dipole radiation
\begin{align}
\vec j_H(\vec x) = \frac{1}{4\pi}\vec \nabla \times \vec B(\vec x) .
\end{align}
Thus all the fields, the vector potential $\vec A_H(\vec x)$, the vortex field
$\vec B(\vec x)= \vec \nabla \times \vec A_H(\vec x)$
and the vortex density $\vec j_H(\vec x)=\vec \nabla \times \vec B(\vec x)/4 \pi$
decay asymptotically as $1/r$.
This is a consequence of retardation.
We also note that the last term in  \eqref{vektorpot}, the regularization term, which guaranties the convergence
for a weak decrease of the field as $1/r$, is not necessary in this case.
The integrals converge even without this term \cite {stewart}.
This also applies  for other fields like $\vec v(\vec x)= \vec p/r$.
On the contrary, for a vector field like
$\vec v(\vec x)= \vec e_r/r$ the regularization term is necessary for reaching convergence,
but the regularization point $\vec x_0$ has to be different from zero.
Then we get for the potential 
\begin{align}
\phi(\vec x)=\ln r_0 - \ln r\, .
\end{align}

One may be surprised that all calculations for the radiation field could be performed with a regularization on a lower level
($G_0$ instead of $G_1$ etc) than expected according to the decay of the vector field.
One reason lies in the symmetries of the sources and circulations (see App. \ref{appc}).
\section{Diverging vector fields }
\subsection{Supplement to the fundamental theorem of vector analysis\label{extended}}

As already mentioned, the fundamental theorem of vector analysis can be applied to asymptotically sublinearly {\bf diverging} vector fields,
if one inserts the faster decaying Green's function $G_2$  \eqref{g2} into  \eqref{skalarpot} and \eqref{vektorpot} 
for the computation of $\phi$ and $\vec A$. 
\par \smallskip \noindent
{\sl Let $\vec v(\vec x)$ be an everywhere continuous differentiable vector field  with the asymptotic behavior
$\displaystyle \lim_{r \to \infty} v(r)/r^{1-\epsilon} < \infty$, with $\epsilon > 0$}, 
{\sl  then the decomposition}
\begin{equation}\label{vektorsatze}
\vec v(\vec x) = \vec v_l + \vec v_t+\vec v_c
= - \vec \nabla \tilde \phi(\vec x) + \vec \nabla \times \vec {\tilde A}(\vec x)+\vec v(\vec x_0)
\end{equation}
{\sl with}
\begin{align} \label{skalarpote} 
\tilde \phi(\vec x) &=   \frac{1}{4\pi}\int d^3 x'\,\big (\vec \nabla' \cdot \vec v(\vec x')\big )
G_2(\vec x-\vec x_0,\vec x'-\vec x_0) \\
\label{vektorpote} 
\vec {\tilde A}(\vec x) &= \frac{1}{4\pi}\int d^3 x'\,(\vec \nabla' \times \vec v(\vec x')) G_2(\vec x-\vec x_0,\vec x'-\vec x_0) 
\end{align}
{\sl is unique apart from a constant vector field.}
\par \smallskip \noindent
\textit{Remarks}:
\begin{itemize}
\item  The vector potential $\vec A$ by its definition is purely transversal, $\vec \nabla \cdot \vec A=0$ (sse proof below).
\item The regularization of the Green's function at a point $\vec x_0$ is responsible for the finiteness of the integrals
 \eqref{skalarpote} and \eqref{vektorpote}.
The point can be chosen arbitrarily. 
\item Curl- and divergence-free (harmonic) fields $\vec v_h$ can be added to $\vec v_l$ if they are
subtracted from $\vec v_t$ without affecting $\vec v$. 
\item Harmonic fields with the exception of constant vector fields $\vec v_c$ can be suppressed
if one demands that $\vec v_l$ asymptotically diverge weaker as linearly.
This is shown in section \ref{unique}.
\item If $\vec v_l$ and $\vec v_t$ are calculated with  \eqref{skalarpote} and \eqref{vektorpote}, 
then one obtains  the value of the constant vector $\vec v_c=\vec v (\vec x_0)$ depending on the arbitrary  regularization point. 
\item  
If the vector field approaches zero slower than any power law or if it diverges logarithmically
(as it is the case in Blumenthal's formulation of the theorem),  or if it increases sublinearly, then    
the faster converging Green's function $G_2$ has to be applied in $\phi$ and $\vec A$.
The price one has to pay for this weaker requirements on the vector field $\vec v$ is the loss of the rigorous uniqueness  of the decomposition. 
\item If one uses the regularized Green's function $G_2$  for the case where the vector field $\vec v$ decreases stronger,
one recovers the unique decomposition of the fundamental theorem \eqref{vektorsatz},
since all integrals coming from the regularization terms are finite.
\item  In the special case of the theorem where $\vec v$ approaches zero at infinity weaker as any
power of $1/r$ (the case $\epsilon=1$), then $v_l$ and $v_t$ may diverge logarithmically although the sum of the two parts decays to zero
\cite{blumenthal}. 
\end{itemize}
\subsubsection{Proof of the supplementary theorem} 
At first one has to show the existence of $\tilde \phi$ and $\vec {\tilde A}$,  \eqref{skalarpote}
and \eqref{vektorpote}.
Concerning the potentials $\tilde \phi$ and $\vec {\tilde A}$, their integrand has the same asymptotic decay
governed by $\vec v(\vec x') G_2(\vec x') \sim 1/r^{2+\epsilon}$, as $\phi$ and $\vec A$ in the former proof for the fundamental theorem.
$G_2(\vec x,\vec x')$  \eqref{g2} has compared to $G_1(\vec x,\vec x')$ an additional singular term at $\vec x_0=0$.
We have to prove that the contribution of this singularity to $\tilde \phi$ (and $\vec {\tilde A}$) is finite.   
For this purpose we integrate over a small sphere of radius $\eta \to 0$ around zero ($\xi' = \cos \vartheta'$) 
\begin{align*}
\tilde \phi_\eta (\vec x)&= \frac{-1}{4\pi} \int_{S_\eta} d^3 x'\,\big (\vec \nabla' \cdot \vec v(\vec x')\big )
\frac{\vec x \cdot \vec x'}{r'^3} 
= -2\pi\rho(0)\, r  \int_0^\eta d r' \int_{-1}^1 d \xi' \, \xi'= 0. 
\end{align*}
Now we can be sure that $\tilde \phi(\vec x)$ and $\vec{\tilde A}(\vec x)$ exist. 
Starting from  \eqref{skalarpot}, we can reformulate all equations up to  \eqref{vlplusvt} by replacing 
$G_i$ by $G_{i+1}$.
\par \medskip \noindent
Before the decomposed vector fields are computed, one should compare the scalar potentials  \eqref{vektorpot} with \eqref{vektorpote}.
Because of the use of $G_2$  in $\tilde \phi$ these both potentials differ in linear function in $\vec x$.
This applies even to  the difference between $\vec A$  and $\vec {\tilde A}$ and has the consequence that
$\vec v_l$ and $\vec v_l$ are indeterminate by a constant vector.   
Starting with  \eqref{skalarpote} we build  $\vec v_l$ as the negative gradient of $\tilde \phi$ and
check if $\vec v_l$ has the same sources a $\vec v$
\begin{align} \label{a.vlneu}
\vec v_l(\vec x) &= - \vec \nabla \tilde \phi(\vec x)= -\frac{1}{4\pi}\int d^3 x'\, (\vec \nabla'\cdot \vec v(\vec x'))
 \vec \nabla G_2(\vec x-\vec x_0,\vec x'-\vec x_0)
\\ \notag
\vec \nabla \cdot \vec v_l(\vec x) &= -\frac{1}{4\pi}\int d^3 x'\,(\vec \nabla'\cdot \vec v(\vec x'))
 \Delta  G_2(\vec x-\vec x_0,\vec x'-\vec x_0)= \vec \nabla \cdot \vec v(\vec x).                                                   
\end{align} 
We find that everything holds as expected.
Now we rewrite $\vec v_l$ by using  \eqref{nabg2} and perfoming a partial integration  
\begin{align} \label{a.vlbneu}
\vec v_l(\vec x) &
= \int d^3 x'\, \rho(\vec x')\,
 \vec \nabla'
 \big ( \frac{1}{|\vec x'-\vec x|} - \frac{1}{|\vec x'-\vec x_0|}\big ). 
\end{align} 
As can be seen from  \eqref{a.vlbneu} one gets $\vec v_l(\vec x_0)=0$.
If one compares $\vec v_l(\vec x)$ computed with $G_0$ in  \eqref{a.vlb} one sees that the  additional 
term of $G_1$ subtracts a (divergent) constant field from the first term to hold $\vec v_l(\vec x)$ finite.  
This divergence- and curl-free vector field does not contribute to the source density.
\par \medskip \noindent
A partial integration in the  vector potential   \eqref{vektorpote} yields to (see  \eqref{vecp}) 
\begin{align}  
\vec {\tilde A}(\vec x) &
=\frac{1}{4\pi}\int
d^3 x'\,\vec v(\vec x')  \times \vec \nabla' G_2(\vec x-\vec x_0,\vec x'-\vec x_0) \, .
\end{align}
As already mentioned, differ$\vec A$ and $\vec {\tilde A}$ by a vector linearly in $\vec x$.
Even $\vec {\tilde A}(\vec x)$ is purely transversal (use   \eqref{nabg2} and compare the result with  \eqref{diva})  
\begin{align} \label{divae}
\vec \nabla \cdot \vec {\tilde A} (\vec x)&= \frac{-1}{4\pi}\int d^3 x'\,
\big(\vec v(\vec x')  \times \vec \nabla' \big)  \cdot \vec \nabla' 
G_1(\vec x-\vec x_0,\vec x'-\vec x_0)
=0  
\end{align}
Now the vortex field is calculated from the vector potential by taking its curl  
and transform $\vec \nabla G_2$ to $- \vec \nabla' G_1$
\begin{align} \label{a.vtneu}
\vec v_t(\vec x)  &= \vec \nabla \times \vec {\tilde A}(\vec x)
= \frac{1}{4\pi}\int d^3 x'\,(\vec \nabla' \times \vec v(\vec x')) \times \vec \nabla' 
G_1(\vec x-\vec x_0,\vec x'-\vec x_0) \, . 
\end{align}
Analogous in the case of the irrotational vector $\vec v_l$ also
 \eqref{a.vtneu} contains in addition to  \eqref{a.vt} a constant vector field effecting that $\vec v_t(\vec x_0)=0$. 
Now it is shown that the vortices of $\vec v_t$ are the same as for $\vec v$.
Inserting $G_2$ into  \eqref{rotvt}  one obtains that $\vec \nabla \times \vec v_t=\vec \nabla \times \vec v$.
\par \medskip 
Our interest is directed to the sum  $\vec v_l + \vec v_t$ since both fields have divergence- and curl-free constant vectors.
They inhibit that $\vec v$ is the sum of $\vec v_l+ \vec v_t$ as the following calculation shows.  
If one replace in  \eqref{vlplusvt} $G_2$ by $G_3$ and takes $\vec x_0 \ne 0$ one
obtains
\begin{align} 
\vec v_l(\vec x)+ \vec v_t(\vec x) &
= \frac{-1}{4\pi}\int d^3 x'\,
  \Delta \big (\vec v(\vec x')\,G_3(\vec x-\vec x_0,\vec x'-\vec x_0)\big )
= \vec v(\vec x)- \vec v(\vec x_0). 
\end{align} 
Now we identify the constant vector $\vec v(\vec x_0)$ with $\vec v_c$ in  \eqref{vektorsatze}. 
Thus $\vec v_c$, $\vec v_{l}$ and $\vec v_t$ depend all on $\vec x_0$.
\subsubsection{On the uniqueness in the case of an increasing vector field\label{unique}}

Let us start from two different solutions of the decomposition
 $\vec v_{l,t}$ and $\vec v'_{l,t}$ with the same sources and vortices respectively.
Then the difference  $\vec v_d  = \vec v_l  - \vec v'_l$ is a irrotational solenoidal vector field.

This vector field can be written as the gradient of a scalar potential $\phi_d$ that fulfills
the Laplace equation  $\Delta \phi_d=0$ in the whole space.
Its most general solution in spherical coordinates reads:
\begin{equation}  \label{g2.laplacepot1}
\phi_d (r, \vartheta, \varphi) = \sum^\infty_{l=0} \sum^l_{m=-l}
(\alpha_{lm} r^l + \beta_{lm} r^{-l-1}) Y_{lm} (\vartheta, \varphi)
\end{equation} 
where $\alpha_{lm}$ and  $\beta_{lm}$ are coefficients which allow  the solution to fulfill the boundary conditions and
$Y_{lm}$ are the spherical harmonics.
All $\beta_{lm}$ vanish because the zero-point is contained within the domain and the solution should be regular.
We now calculate the radial harmonic flux of the vector field
\begin{align} \label{g2.flux}
\vec v_d  \cdot \vec e_r &= v_{dr}= - \frac{\partial \phi_d}{\partial r}
= - \sum^\infty_{l=1} \sum^l_{m=-l}
\alpha_{lm}\, l\, r^{l-1}\, Y_{lm} (\vartheta, \varphi). 
\end{align} 
When the distance $r$ goes to $\infty$ and one notes that $v_d \sim r^{1- \epsilon}$ with  $\epsilon  > 0$,
then  all coefficients with $l  -1  >1  - \epsilon$ have to vanish, otherwise this
would lead to a stronger divergence of $v_{dr}$.
Thus only the terms with $l  = 0$ und $l  =1$ remain.  
Therefore the solution reads: 
\begin{equation} 
\phi_d (r, \vartheta, \varphi) = \alpha_{00} Y_{00} +
\sum_{m=-1}^1 \alpha_{1m}Y_{1m}(\vartheta,\varphi)\,r 
= \frac{\alpha_{00}}{\sqrt{4 \pi}} - \vec w  \cdot \vec x  \, ,
\end{equation}   
and we obtain
\begin{align}
\vec v_d(\vec x) &= - \vec \nabla  \phi_{d} = \vec w  \, .
\end{align}   
Hence the field $\vec v_d = \vec w$ is unique up to a constant vector.
The choice of the regularization point influences the constant vector of $\vec v_l$ only.

\subsubsection{Comment on the application of the supplementary theorem} 

We are not aware of a an analytically calculable physical example in this case, however for numerical calculations the knowledge of the validity of such a theorem is important.
There are cases where one does not know always the exact asymptotic behavior of a vector field.
We have already seen that in the case of electromagnetic radiation because of the (symmetry) properties of the field
one could decompose the field with the Green function $G_0$ although generally $G_1$ should be necessary.
A few statements on the influence of the symmetry are made in the appendix \ref{appc}.

Generally we want to point out that the application of the regularization schema  is not only restricted to the decomposition of 
a vector field, but may also be  important for other problems where the solution of a poisson equation (scalar or vectorial) is used.

If one considers  sources that remain finite at infinity,  the vector field belonging to this sources should diverge linearly or stronger.
Within our schema one then has tu use the next step in the regularization procedure.
There the vector field has to be computed with $G_2$ and is determined only up to a  vector field linear in $\vec x$.

\subsubsection{Stronger diverging vector fields\label{strongdiverging}}
We have already seen that asymptotically strong decaying  Green's functions $G_i$ with $i \ge 3$ cannot
be treated in the same manner a those for $i \le 2$.
We restrict ourselves to $i=3$ what means that the vector field may increase less than quadratically.
In this case the potentials are  given by (see  \eqref{phi3})
\begin{align} \label{pot3}
\bar \phi_3(\vec x)&=\int d^3 x'\,\rho(\vec x') G_3(\vec x - \vec x_0,\vec x'-\vec x_0)
+ 2\pi \frac{\rho(\vec x_0)}{3} \,r^2
\\ \notag
\vec {\bar A}_3(\vec x)&=\int d^3 x'\,\vec j(\vec x') G_3(\vec x - \vec x_0,\vec x'-\vec x_0)
+ 2\pi \frac{\vec j(\vec x_0)}{3} \,r^2. 
\end{align}
The last term  of $\phi_3$ and $\vec A_3$ cancels the contribution to the inhomogeneity caused by $G_3$.
Now, $\vec v$ can again be decomposed in $\vec v_l$ and $\vec v_t$ except for a linear vector field that depends on
the regularization point $\vec x_0$.  

\par \smallskip \noindent 
\textit{Remarks}:
\begin{itemize}
\item Besides our statements to the regularization of weak diverging vector fields 
high symmetric vector fields can be decomposed by the method shown here even if they diverge stronger than assumed so far.
This can be seen from  \eqref{si} for $i=0$ (and $i=1$), where the additional terms to $G_1$ in $G_2$ cancel
the contributions to $S_0$  (and $S_1$) for $r' > r$. 
\item Simple examples for this feature are $\vec v = \vec a r^\alpha$ and $\alpha > -1$ or $\vec v = \vec e_r r^\alpha$.
\end{itemize}  
\begin{table}
\newlength{\Wl}
\settowidth{\Wl}{asymptotic}
\begin{tabular}{|c|c|c|c|c|} \hline
\parbox{\Wl}{asymptotic \\[-1.8ex]region\rule[-1mm]{0mm}{0mm}}&Exponent&%
\parbox{\Wl}{Green's \\[-1.8ex]function\rule[-1mm]{0mm}{0mm}}&
$v_l(r\to \infty)$&Unique up to\\ \hline \hline

$v\sim r^{1-\epsilon}$ & $ 0 <\epsilon \le 1$& $G_2(\vec x, \vec x')$ &
$v_l\sim r^{1-\epsilon}$ & $\vec w$ \\ \hline

$v\sim 1/r^\epsilon $  & $0<\epsilon \le 1$& $G_1(\vec x, \vec x')$&
$v_l(\infty)=0$&$\vec w=0$ \\ \hline
            
$v\sim 1/r^{1+\epsilon}$&$0<\epsilon$ & $G_0(\vec x, \vec x') $ &
$v_l(\infty)=0$&$\vec w=0$ \\ \hline
\end{tabular}  
\caption{\label{tab} Different cases of a vector fields $\vec v(\vec x)$, which decay asymptotically to zero
or increase sublinearily (first and second column), can be decomposed into longitudinal (irrotational, curl-free)
$\vec v_l$ and transversal (solenoidal, divergence-free) parts besides a constant vector field $\vec w$.
In order to cover all cases one has to introduce regularized Green's functions (see  \eqref{g1}) (third column)
respectively.
Also shown is the asymptotic condition on the longitudinal field (fourth column).
We also indicate the extent of the uniqueness of the decomposition (fifth column).}                                              
\end{table}

\section{Conclusion}

We have presented a  proof of the fundamental theorem of vector analysis (Helmholtz' decomposition theorem)
for vector fields decaying weakly  and extended to even sublinearly diverging vector fields.
Contrary to the original proof\cite{blumenthal} we can  distinguish between different cases.
Our results are summarized in Tab. \ref{tab}. Note however that not only the decay of the vector field is important
for introducing a regularization but also its symmetry.
This extends the presentations of this theorem given usually in textbook on electrodynamics.
Especially the case of weakly decaying fields has been discussed in the physical literature in the context
of electromagnetic radiation fields.

Considering the validity of Helmholtz' decomposition theorem there is no doubt that the
theorem can be applied quite generally to electromagnetic fields either static or dynamic.
This was demonstrated by explicit examples. 
\begin{acknowledgments}
One of the authors (D. P.) thanks W. Zulehner for helpful discussions.
\end{acknowledgments}     
\begin{appendix}

\section{Existence of the vector potential and its transverse vector field \label{vec25}}
For the vector potential   \eqref{vektorpot} we have the problem that we do not to know the asymptotic
behavior of the vortex density.
Therefore we show the existence of the integral \eqref{vektorpot} in the same way as in subsection \ref{phi21} 
and obtain after a partial integration  
\begin{align} \label{vecp} 
\vec A(\vec x) &
= \frac{1}{4\pi}\int d^3 x'\,\vec v(\vec x')  \times \vec \nabla' 
G_1(\vec x,\vec x'). 
\end{align}
This vector potential $\vec A(\vec x)$ turns out to be a purely tranversal vector potential  for which the divergence vanishes.
In order to show this we use  \eqref{nabg2} in  \eqref{vecp}
\begin{align} \label{diva}
\vec \nabla \cdot \vec A (\vec x)&= \frac{-1}{4\pi}\int d^3 x'\,
( \vec v(\vec x')\times \vec \nabla' )  \cdot \vec \nabla' G_0(\vec x,\vec x')
=0. 
\end{align}
 
The fundamental theorem of vector analysis states that the solenoidal part of $\vec v$ field  is given by (see  \eqref{vecp})
\begin{align}\label{a.vt}
\vec v_t(\vec x)&=\vec \nabla \times \vec A(\vec x)
= \frac{1}{4\pi}\int d^3 x'\, \vec \nabla \times\big (\vec \nabla' \times \vec v(\vec x')\big ) G_1(\vec x,\vec x')
\, . 
\end{align}
Since $\vec v_t$ is calculated from the vector potential div\,$\vec v_t$ is zero.

Now one has to show that the vortices of $\vec v_t$  are the same as those of the given vector field $\vec v$.
Therefore we calculate the curl of   \eqref{a.vt} and use the identity 
$(\vec \nabla \times \vec \nabla) \vec v = \vec \nabla (\vec \nabla \cdot \vec v)  - \Delta  \vec v$
\begin{align} \label{rotvt}
\vec \nabla \times \vec v_t(\vec x)&
= \frac{1}{4\pi}\int d^3 x'\,\vec \nabla  \times \big [\vec \nabla \times
\big (\vec \nabla' \times \vec v(\vec x')\big ) G_1(\vec x,\vec x') \big ] 
\notag \\ &
= \frac{-1}{4\pi}\int d^3 x'\,\Delta G_1(\vec x,\vec x')
\big (\vec \nabla' \times \vec v(\vec x')\big )
= \vec \nabla \times \vec v(\vec x). 
\end{align}
$\vec v_t$ is a pure solenoidal field and its 
vortices of $\vec v_t$ are the same as those of $\vec v$. 
\par \medskip  
In a final step it is shown the sum of the irrotational and solenoidal field  $\vec v_l+ \vec v_t = \vec v$
equals the given vector field.   
For this reason we reshape $\phi(\vec x)$  \eqref{phi} and $\vec A(\vec x)$  \eqref{vecp} by replacing $\vec \nabla' G_1(\vec x,\vec x')$ 
with $-\vec \nabla G_2(\vec x,\vec x')$ according to  \eqref{nabg2}
\begin{align} \label{phig2}
\phi(\vec x)  & =
 \frac{1}{4\pi} \int  d^3 x'\,\vec \nabla \cdot  \vec v(\vec x')\,G_2(\vec x,\vec x') 
\\ \notag
\vec A(\vec x) &
= \frac{1}{4\pi}\int d^3 x'\, \vec \nabla \times \vec v(\vec x')  \ 
G_2(\vec x,\vec x'). 
\end{align}  
Now the negative gradient of $\phi(\vec x)$ is added to the curl of $\vec A(\vec x)$ and the identity 
$(\vec \nabla \times \vec \nabla) \vec v = \vec \nabla (\vec \nabla \cdot \vec v)  - \Delta  \vec v$ is used 
\begin{align} \label{vlplusvt}
\vec v_l(\vec x)+ \vec v_t(\vec x) &= \frac{-1}{4\pi}\int d^3 x'\,
 \Big [ \vec \nabla (\vec \nabla \cdot \vec v(\vec x')) -  \vec \nabla\times( \vec \nabla \times \vec v(\vec x'))\Big ]
      G_2(\vec x, \vec x') 
\notag \\ & 
= \frac{-1}{4\pi}\int d^3 x'\,
  \Delta \big (\vec v(\vec x')\,G_2(\vec x,\vec x')\big )= \vec v(\vec x). 
\end{align}

\section{Example of a diverging vector field \label{appb}}
We want to study the following vector field
\begin{align} \label{divergingfield} 
\vec v &= \vec a \times ( e_r \times \vec a)\sqrt{r} 
\end{align}
where $\vec a$ is  a constant vector. $\vec v$ diverges as $\sim \sqrt{r}$.
It seems to be more convenient to determine first sources and vortices 
and then to calculate the fields belonging to these 
\begin{align}
\rho (\vec x) &
=\frac{1}{4\pi}\big [3 a^2 +  (\vec a \cdot \vec e_r)^2 \big ]\frac{1}{2 \sqrt{r}}
&  
\vec j(\vec x) &
=\frac{1}{4\pi}(\vec a  \cdot \vec e_r) \vec e_r \times \vec a \frac{1}{2 \sqrt{r}}.   
\end{align}
To make the computation of the potentials as simple as possible we use the regularization point $\vec x_0=0$. 
Then we get $\phi$ from  \eqref{skalarpote} es follows 
\begin{align} \label{bsp.phi}
\phi(\vec x) &= \frac{1}{4\pi} \int d^3 x'\,\frac{1}{2 \sqrt{r'}}
\big [3a^2 + (\vec a \cdot \vec e_{r'})^2\big ] G_2(\vec x,\vec x')
\end{align}
In the next step we introduce spherical coordinates and 
we fix the primed coordinate system by the unprimed vector $\vec x$:
$\vec e_{z'} = \vec e_r$ and  
perform the integration over the azimuth $\varphi'$  ($\xi'=\cos \vartheta'$). 
This leads to a replacement of  $\sin \varphi'\, \cos \varphi'$ by zero and $\cos^2 \varphi' = \sin^2 \varphi'$ by $1/2$ 
\begin{align*}
\vec a  \cdot \vec e_{r'} &= a_{x'} \sqrt{1 - \xi'^2}\cos \varphi' 
                              + a_{y'}\sqrt{1-\xi'^2}\sin \varphi'  + a_{z'} \xi'    
\\
(\vec a  \cdot \vec e_{r'})^2 &
= a^2\frac{1}{2}(1 - \xi'^2) + (\vec a \cdot \vec e_r)^2 \frac{1}{2}(3 \xi'^2 -1) 
\end{align*}
For the calculation of the angular integral of $\phi$ one needs to evaluate this two surface integrals
($ d\Omega' = d \xi'\ d\varphi'$)
\begin{align} \label{si}
S_i(r,r') &=\frac{1}{4\pi}\int d\Omega' \,\xi'^i\,G_2(\vec x,\vec x')
\\ 
S_0(r,r') &=\big (\frac{1}{r}-\frac{1}{r'}\big )\theta(r-r') 
\\
S_2(r,r') & = \frac{1}{3} S_0 + \frac{2}{15} \big \{  \frac{r'^2}{r^3} \theta(r  -r') 
 +  \frac{r^2}{r'^3} \theta(r' - r)\big \} 
\end{align}
Now we get for the scalar potential
\begin{align} \label{phidiverge}
\phi(\vec x) &= \int_0^\infty d r'\,\sqrt{r'}^3\frac{1}{4}\Big \{\big [7 a^2 -(\vec a\cdot \vec e_r)^2\big ]S_0(r,r')
-\big[a^2 - 3 (\vec a\cdot \vec e_r)^2\big ] S_2(r,r')\Big \} \notag 
\\  
&= \frac{1}{9}\big [ -7a^2 + 2 (\vec a\cdot \vec e_r)^2\big ]\sqrt{r}^3 
\end{align}
The analogous calculation for the vector potential yields 
\begin{align} 
\vec A(\vec x) &
= \frac{2}{9} \vec a \times (\vec e_r \times \vec a) \sqrt{r}^3.
\end{align}
In   the last step, the calculation of the decomposed vector fields,
we get    
\begin{align}
\vec v_l(\vec x) &= - \vec \nabla \phi(\vec x) 
 = - \frac{1}{9}\Big \{\big [7a^2 - (\vec a\cdot \vec e_r)^2\big]\vec e_r
 +4(\vec a \cdot \vec e_r)^2 \vec e_r \Big \}\sqrt{r}
\\
\vec v_t(\vec x) &=  \vec \nabla \times \vec A(\vec x) 
= \frac{1}{9}\Big \{2 \vec a \times( \vec e_r \times \vec a)
- 4 (\vec a\cdot \vec e_r) \vec a 
+ (\vec a\cdot \vec e_r)\vec e_r \times (\vec a \times \vec e_r)\Big \}\sqrt{r}.
\end{align}
Thus we have demonstrated that sublinearly divergent vector fields can be decomposed in 
its irrotational and solenoidal components, both diverging as $\sim \sqrt{r}$. 
Since $\vec v(\vec x_0=0)$ vanishes, it is indeed $\vec v_l+\vec v_t=\vec v$.

\section{On the influence of symmetry \label{appc}} 

In the preceding example we used \eqref{si} for the computation of $\phi(\vec x)$. 
The $\theta$-functions in \eqref {si} cuts all diverging contributions for $r'\to \infty$  in  \eqref{phidiverge}.
Similarily a Helmholtz vector  potential $\vec A(\vec x)$ calculated  for a circulation  with the symmetry  $\vec j(\vec x)= f(r)\vec p $ exists for arbitrary diverging $f(r)$.

In the case of the electromagnetic example in Sec. \ref{rad} the Helmholtz vector potential  $\vec A_H(\vec x)$ of the electric radiation field  \eqref{ahele} the same effect happens.
The required surface integral is calculated for a  current \eqref{bfield} of the form $\vec j_H=f(r)\,\vec e_r \times \vec p$ and reads
\begin{align*}
S_1^0(r,r')=
\frac{1}{4\pi}\int  d \Omega' \,\xi'\frac{\xi'}{|\vec x - \vec x'|}  
 \frac{1}{3}\big [\frac{ r'}{r^2} \theta(r -r') + \frac{r}{r'^2}\theta(r' -r)\big ].  
\end{align*} 
The integral converges for $r'\to \infty$ with $1/r'^2$ which is  stronger  as expected  from the behavior of $G_0$ alone,
but one gets  the same result for the surface integral if one uses  $G_1$ instead of $G_0$.
This makes clear that the type of regularization necessary depends on the symmetry of the vector field $\vec v$ considered
and why we did not need $G_1$ in the case of electromagnetic radiation.
\end{appendix}
  

\begin{thebibliography}{99}
\bibitem{strahlung} There were several items to clarify for time dependent vector fields, especially the question of  retardation, its connection to causality and the choice of gauge.
\bibitem{altzit} see \cite{petra}  (8.4.5) and (8.4.6) p. 294. The time dependence is in this case contained in the Fourier factor $e^{-i\omega t}$
\bibitem{rho} The factor $e^{i k r}(1- i k r)$ is a consequence of the singular structure of $\rho_p$ and can be replaced by 1 in all calculations.
\bibitem{komm} The appearance of the quasistatic potentials has led to a discussion on the validity of causality
in the case of Coulomb gauge.
However it was recognized earlier \cite{brill} and confirmed later \cite{rohrlich}, that the physical quantities are causal
and the decomposition is valid also for time dependent (retarded) fields.
\end{thebibliography}
\end{document}